\documentclass[10pt,conference]{IEEEtran}

\usepackage{cite}
\usepackage{amsmath,amssymb}
\usepackage{graphicx}
\usepackage{url}
\usepackage{array}
\usepackage{multirow}
\usepackage{multicol}
\usepackage{listings}
\usepackage{xcolor}
\usepackage{threeparttable}
\usepackage{makecell}
\usepackage{booktabs}
\usepackage{tabularx}
\usepackage{xurl}
\usepackage{hyperref}
\usepackage{adjustbox}
\usepackage{orcidlink}
\usepackage{fontawesome}

\usepackage[most]{tcolorbox}

\begin{document}

\title{An Empirical Study on Improving LLM-Based Issue Classification through Domain-Informed Analysis}

\author{
\IEEEauthorblockN{
Anubhav Gupta\,\orcidlink{0000-0003-0187-8192}
}
\IEEEauthorblockA{
University of British Columbia \\ 
Kelowna, BC, Canada \\
anubhav.gupta@ubc.ca
}
\and
\IEEEauthorblockN{
Tanner Wright\,\orcidlink{0009-0003-2267-1622}
}
\IEEEauthorblockA{
University of British Columbia \\
Kelowna, BC, Canada \\
twrigh01@student.ubc.ca
}
\and
\IEEEauthorblockN{
Gema Rodr\'iguez-P\'erez\,\orcidlink{0000-0002-0062-8418}
}
\IEEEauthorblockA{
University of British Columbia \\ 
Kelowna, BC, Canada \\
gema.rodriguezperez@ubc.ca
}
}

\maketitle

\begin{abstract}
Issue reports that describe Intrinsic Bugs, Extrinsic Bugs, and Not a Bug cases often use similar language, but they might point to different causes and require different maintenance actions. This makes three-way issue classification difficult for Large Language Models (LLMs), particularly when prompt-based refinements begin to plateau. We study whether domain-informed analysis of LLM outputs can improve classification beyond prompt modification alone. Using 428 manually labeled issue reports, we evaluate six LLMs across 18 prompt configurations and analyze 7,704 model-generated rationales to identify recurring failure patterns in how models justify predictions, use evidence, and handle conflicting information. We translate these findings into a multi-stage pipeline that separates bug validity from bug origin. The pipeline improves macro F1 and Extrinsic Bug recall for five of the six models. The strongest pipeline configuration reaches 0.827 accuracy and 0.757 Extrinsic Bug recall, improving over the best-accuracy prompt-only baseline by +1.2 and +7.5 pp, respectively. The decomposition further reveals that bug origin classification is more reliable than validity, indicating that the bug-validity decision accounts for most of the remaining error. These results show that expert failure analysis can support more reliable LLM-based software engineering classifiers.

\end{abstract}


\section{Introduction}

Issue reports are central artifacts in software maintenance, but their interpretation is often more difficult than it first appears. A report may mention crashes, failing tests, stack traces, unexpected behavior, or error messages and still not describe a software defect. Conversely, a report may describe a real defect whose root cause lies outside the focal project's own source history, for example in a changed dependency, platform configuration, external service, or upstream API. This makes issue classification a compound problem: it requires deciding both \emph{bug validity}~\cite{fan2018chaff}, i.e., whether the report describes a real defect, and \emph{bug origin}~\cite{rodriguez2020how}, i.e., whether the root cause of a valid defect is intrinsic or extrinsic to the focal project. In this paper, we study this problem as a three-way classification task: \textit{Intrinsic Bug}, \textit{Extrinsic Bug}, or \textit{Not a Bug}.

This combined validity-origin distinction matters for both maintenance practice and empirical software engineering research. Non-bug reports may require documentation updates, clarification, duplicate handling, or enhancement triage, while valid bug reports require root-cause analysis. Treating non-bugs or extrinsic bugs as intrinsic project defects can contaminate defect repositories, misdirect root-cause analysis, and distort conclusions drawn from software analytics studies~\cite{herzig2013feature,wu2021data,zhang2025less}. This problem is not marginal: Herzig et al. manually inspected more than 7,000 issue reports and found that 33.8\% of reports labeled as bugs did not describe corrective maintenance~\cite{herzig2013feature}. Prior work has also shown that intrinsic and extrinsic bugs can differ in ways that affect downstream defect prediction~\cite{rodriguez2020how,rodriguez2020watch}. As software engineering increasingly adopts automated classification and decision-support systems, the quality of these distinctions becomes a prerequisite for trustworthy automation.

However, this three-way validity and origin classification task is not a simple text-classification problem. A report that mentions a dependency is not necessarily an \textit{Extrinsic Bug}; the dependency may only appear in the execution context, or the focal project may be using the dependency incorrectly. Similarly, a report that uses failure-like language is not necessarily a bug; it may describe expected behavior, user error, an unsupported configuration, a documentation problem, or a feature request. Correct classification often depends on domain knowledge, expected behavior, project responsibility, and evidence distributed across the issue discussion.

Large Language Models (LLMs) are a natural candidate for this task because they can classify issue text from instructions and examples without requiring a task-specific classifier to be trained from scratch~\cite{brown2020language,hou2024large}. Prior studies show that GPT-like and transformer-like models can automate issue labeling~\cite{colavito2024leveraging,colavito2025benchmarking}. However, performance metrics alone do not show whether a model has learned the domain distinctions that define the classification problem. Two models may obtain similar F1 scores while failing for different reasons: one may confuse non-bugs with intrinsic bugs, another may overuse the extrinsic label whenever a dependency is mentioned, and another may rely on superficial failure terminology rather than root-cause evidence.  

In this paper, we investigate whether domain expertise can improve LLM-based issue classification beyond prompt-only classification. We evaluate six LLMs under 18 prompt-only configurations on 428 issue reports manually labeled as \textit{Intrinsic Bug}, \textit{Extrinsic Bug}, or \textit{Not a Bug}. We then analyze both model predictions and model-generated rationales to understand how the models appear to interpret the task, which evidence they emphasize, and where their interpretations diverge from expert judgment. Finally, we use this failure analysis to redesign the classification process as a domain-informed multi-stage pipeline. We evaluate the resulting pipeline using class-level precision, recall, and F1, together with macro-averaged performance. 
 
Our results show that prompt-only changes reach an observed ceiling of 81.5\% accuracy and 0.801 macro F1, motivating a deeper analysis of model-output failure patterns. Applying a structured rationale-inspection template to 7,704 representative model-generated rationales reveals two recurring patterns that distinguish correct from incorrect predictions across all three classes: \textit{contradictory evidence mishandling (L2-Q3)} and \textit{conclusion-first reasoning (L2-Q5)}. \textit{Extrinsic Bug} issues emerge as a distinct source of difficulty, exhibiting the highest misclassification rate (50.6\%) and revealing that models frequently revise away from the correct label after initially identifying it. Translating these findings into a domain-informed multi-stage pipeline that separates bug validity from bug origin improves macro F1 for five of six models and raises the best configuration's accuracy from 0.815 to 0.827 (+1.2 pp) while lifting \textit{Extrinsic Bug} recall to 0.757 (+7.5 pp over the best-accuracy baseline). The pipeline also shows that each decision is independently reliable, with bug origin classification (up to 0.907 accuracy) more accurate than bug validity (up to 0.871), suggesting that the validity decision, rather than origin, accounts for much of the remaining error.

This paper makes the following contributions:
\begin{enumerate}
    \item \textit{An empirical evaluation:} We evaluate six LLMs across three prompts (18 model-prompt configurations) for three-way issue classification and analyze performance variability and class-specific errors.
    \item \textit{Rationale Analysis:} We introduce a structured template for inspecting model rationales, focusing on evidence use, rule application, alternative labels, and contradictory information.
    \item \textit{A domain-informed pipeline:} We propose a multi-stage classification process that uses expert failure analysis to improve LLM-based issue classification.
\end{enumerate}

\section{Background and Related Work}
\label{sec:relatedworks}
\noindent\textbf{Bug Origins: Intrinsic and Extrinsic Bugs:} Some software analytics techniques, especially defect prediction and bug-mining approaches, assume that reported failures originate from changes in the same repository~\cite{karampatsis2020often,liu2021automatically,hirsch2022using}. Rodr\'iguez-P\'erez et al.~\cite{rodriguez2020how} challenged this assumption by distinguishing bugs according to their origin. An \emph{intrinsic bug} is caused by a change within the analyzed project's source code history, whereas an \emph{extrinsic bug} is a genuine failure whose cause lies outside that history, for example in a dependency, external service, or upstream component. In a manual study of OpenStack Nova and Elasticsearch, they found that 9--21\% of bugs were extrinsic~\cite{rodriguez2020how}. A follow-up study showed that excluding Extrinsic bugs could change defect-prediction performance by as much as 16 AUC pp~\cite{rodriguez2020watch}. 

Recent work has begun to automate bug-origin classification:  
BuggIn is a TF--IDF-based classifier for identifying Intrinsic bugs from issue reports~\cite{bhandari2023buggin}, and Fan et al.~\cite{fan2026AutomaticExtrinsic} later proposed a TextCNN--KAN model for identifying Extrinsic bug reports. These studies show that textual signals can support origin classification, but they focus mainly on binary classification in a limited ecosystem setting. Wright et al.~\cite{msr2026} extended bug-origin analysis to the npm ecosystem through the InEx-Bug dataset, which manually labels GitHub issues as \textit{Intrinsic Bug}, \textit{Extrinsic Bug}, \textit{Not a Bug}, or \textit{Unknown}. 

\noindent\textbf{Dependency Bugs in Package Ecosystems:} The npm ecosystem is especially relevant for studying \textit{Extrinsic Bug} reports because packages depend on deep and evolving dependency networks. Prior ecosystem studies show that dependency networks grow and reorganize over time~\cite{kikas2017structure,decan2019empirical}, and that downstream packages can be affected by vulnerabilities, technical lag, deprecations, breaking changes, and abandoned dependencies~\cite{decan2018impact,decan2018evolution,cogo2021deprecation,venturini2023depended,miller2025understanding}. 

These ecosystem phenomena are related but not equivalent to Extrinsic Bug. A vulnerability, deprecation, or breaking release is an upstream event or exposure mechanism; an Extrinsic Bug is the downstream failure attributed to such an event. Network studies can estimate how many packages are exposed to external risks, but they generally do not determine whether individual downstream reports describe a causally related failure. The two literatures operate at complementary levels: ecosystem studies explain how external risk can spread, while bug-origin studies identify how such failures are reported, interpreted, and handled in downstream projects.

\noindent\textbf{Automated Classification of Issue Reports: }
Issue-report classification has traditionally focused on validity or type, such as distinguishing bugs from feature requests, questions, documentation issues, or other maintenance tasks~\cite{antoniol2008bug,qin2018classifying,kallis2019ticket,kallis2021predicting}. Herzig et al.~\cite{herzig2013feature} manually examined more than 7,000 issue reports and found that over 30\% of reports labeled as bugs were misclassified. Subsequent work treated bug-versus-non-bug identification as a text-classification problem using bag-of-words, n-grams, TF--IDF, topic models, and classifiers such as logistic regression, support-vector machines, decision trees, and random forests~\cite{antoniol2008bug,Terdchanakul2017,Aktas2023,qin2018classifying,Zhou2014,herbold2020feasibility}. Recent work uses pre-trained language models and LLMs for issue classification~\cite{aracena2024applying}. Colavito et al.~\cite{colavito2024impact,colavito2024leveraging} evaluated transformer-based models for issue classification and showed that data quality and label reliability materially affect evaluation. Koyuncu~\cite{koyuncu2025exploring} studied how model and prompt choices affect fine-grained bug-report categorization. A recent mapping study identified 46 primary studies on automatic bug/non-bug classification and found that traditional machine learning, deep learning, BERT-family models, and zero- or few-shot methods are all represented in the literature~\cite{laiq2026automatic}. LLM-BRC~\cite{du2024llm} further illustrates the use of language-model representations in hierarchical bug classification, although its goal is to classify behavioral properties of bug reports rather than bug origin.

This literature provides useful methods and evaluation practices, but its target labels are not interchangeable with \textit{Intrinsic Bug} and \textit{Extrinsic Bug} labels. A classifier that separates bugs from non-bugs does not determine whether a valid bug was caused inside the focal repository or by an external dependency, platform, or environment. Conversely, bug-origin classifiers such as BuggIn~\cite{bhandari2023buggin} and Fan et al.'s model~\cite{fan2026AutomaticExtrinsic} address origin classification, but do not fully address the preceding validity problem.

\noindent\textbf{Research Gap.}
Prior work supports parts of our problem but not their combination. Bug origin studies show that intrinsic and extrinsic bugs are empirically distinct and can affect software analytics results~\cite{rodriguez2020how,rodriguez2020watch}. Ecosystem studies show that dependency-intensive environments such as npm create many opportunities for externally caused failures~\cite{kikas2017structure,decan2019empirical}
. However, what remains underexplored is how  LLMs fail when bug validity and bug origin must be decided together, and how expert analysis of those failures can be converted into a reproducible improvement process. Our study addresses this gap by evaluating prompt-only LLM classifiers, diagnosing their failure patterns through expert analysis of predictions and rationales, and translating those findings into a domain-informed pipeline for classifying issue reports as \textit{Intrinsic Bug}, \textit{Extrinsic Bug}, or \textit{Not a Bug}. 

\section{Study Design}
\label{sec:studydesign}
Our goal is to investigate \textit{whether} and \textit{how} domain expertise can improve LLM-based classification of software issue reports. We focus on a three-way classification task in which each issue report must be labeled as \textit{Intrinsic Bug}, \textit{Extrinsic Bug}, or \textit{Not a Bug}. In this study, domain experts are researchers or practitioners with experience in bug analysis and bug-origin classification. To address our goal, this paper formulates the following research questions (RQs):

\noindent\textbf{RQ1: How accurately do prompt-only LLM configurations classify issue reports as \textit{Intrinsic Bug}, \textit{Extrinsic Bug}, or \textit{Not a Bug}?}

\noindent\textbf{Motivation.}
Before introducing a domain-informed pipeline, we first establish the performance of prompt-only classification. This provides the baseline against which later pipeline changes are evaluated. It also shows whether prompt effects are consistent across models, prompts, and classes, and identifies which errors persist despite prompt refinement.

\noindent\textbf{RQ2: What recurring rationale-level failure patterns are associated with incorrect LLM classifications, and how do these patterns vary across \textit{Intrinsic Bug}, \textit{Extrinsic Bug}, and \textit{Not a Bug} categories?}

\noindent\textbf{Motivation.}
Aggregate metrics can show where models fail, but they do not reveal how models justify their predictions. RQ2 therefore analyzes model-generated rationales as diagnostic artifacts. The goal is to identify recurring rationale-level patterns that distinguish correct from incorrect predictions and to examine whether those patterns differ across the three ground-truth classes. These patterns provide the empirical basis for the pipeline evaluated in RQ3.

\noindent\textbf{RQ3: Does a domain-informed multi-stage classification pipeline improve performance beyond the strongest prompt-only configuration?}

\noindent\textbf{Motivation.}
Prompt-only classification treats the task as a single three-way decision, but the labels combine two different judgments: bug validity and bug origin. This can lead to errors when the model reasons about bug origin before establishing bug validity, or when it dismisses an extrinsic bug because the focal project is not responsible for the underlying cause. RQ3 tests whether these errors can be reduced by restructuring the classification pipeline itself.

\begin{figure*}[t]
  \centering
  \includegraphics[width=\textwidth]{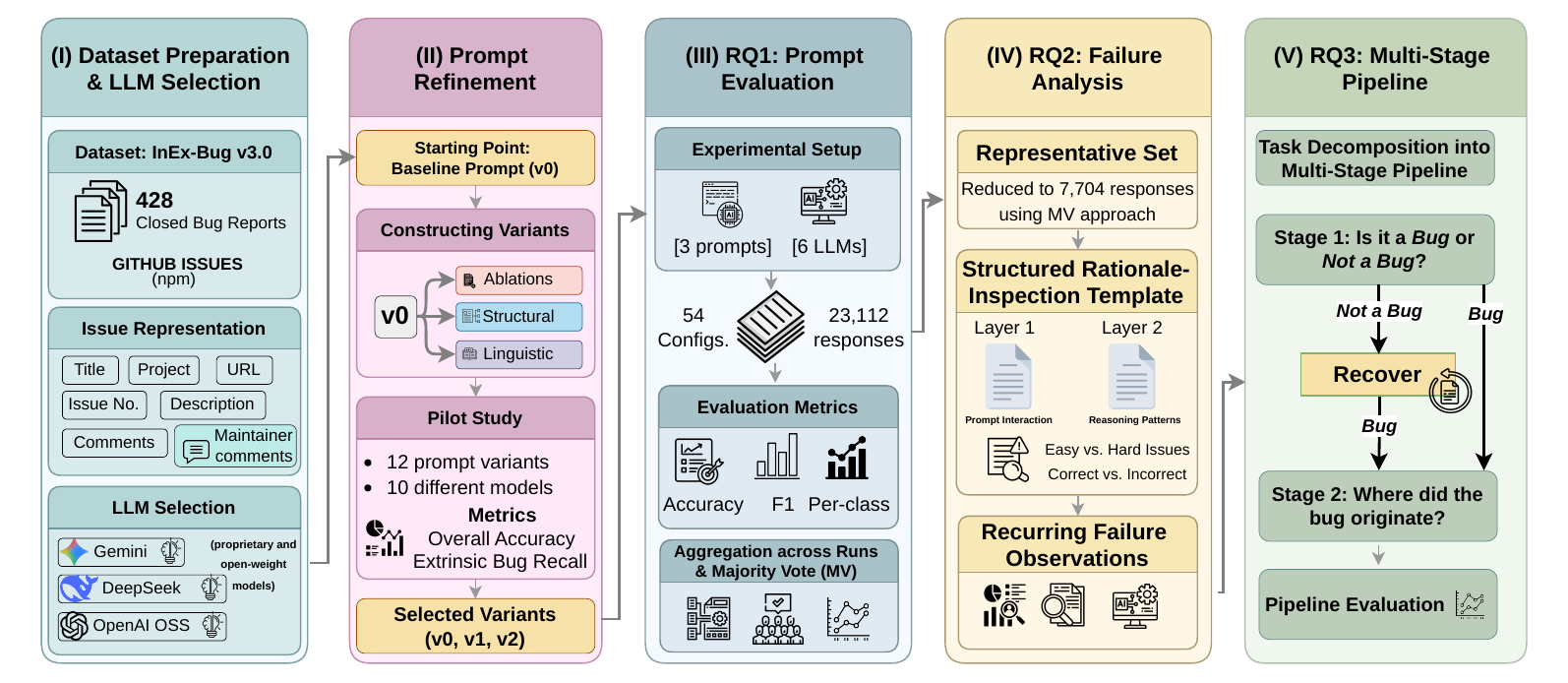}
  \caption{Integrated framework illustrating the domain-informed methodology for LLM-based issue classification.}
  \label{fig:StudyDesignFramework}
\end{figure*}

\section{Methodology}

Figure~\ref{fig:StudyDesignFramework} illustrates the study's five phases: (I) dataset preparation and LLM selection; (II) prompt refinement; (III) prompt evaluation; (IV) qualitative analysis of recurring failure patterns; and (V) construction and evaluation of the domain-informed classification pipeline.

\subsection{Dataset preparation and LLM selection}
\label{sec:dataset}
\noindent\textbf{Dataset:} We use Version~3.0 of the InEx-Bug dataset~\cite{wright2025inexbug}, a manually annotated collection of closed GitHub issue reports from 103 npm repositories. Version~3.0 extends the original dataset~\cite{msr2026} by excluding open and \textit{Unknown} reports and adding 85 newly annotated reports. It contains 428 reports: 214 \textit{Not a Bug}, 107 \textit{Intrinsic Bug}, and 107 \textit{Extrinsic Bug}. The dataset is therefore balanced between bug and non-bug reports, and also balanced between intrinsic and extrinsic bugs within the bug subset. Each record includes issue text (i.e., title, body, labels, and comments), repository and author metadata, timestamps, participants, maintainer activity, reopen events, assignees, milestones, closing pull requests/commits, and the final classification.

\noindent\textbf{Issue Report Representation:} Each issue report is represented as a structured text document that is provided to the LLM as input. Each input includes: (i) the issue \textit{title}; (ii) the \textit{repository name} (e.g., \texttt{axios/axios}); (iii) the \textit{issue number} and corresponding \textit{URL}; (iv) the issue \textit{description}, corresponding to the original report submitted by the user; and (v) the full sequence of \textit{comments} associated with the issue, in chronological order. We also separately extract and append a \textit{maintainer comments} section. Maintainer comments are defined as comments written by users whose GitHub role is marked as \texttt{OWNER}, \texttt{MEMBER}, or \texttt{COLLABORATOR}. We intentionally repeat them in a dedicated section to make maintainer-provided comments more noticeable to the model. This design choice is motivated by the fact that maintainer comments often contain the most reliable evidence for distinguishing between \textit{Intrinsic Bug}, \textit{Extrinsic Bug}, and \textit{Not a Bug} categories, as maintainers are typically the ones who understand the project requirements and boundaries.

\noindent\textbf{LLM Selection and Configuration:} To identify suitable models for our analysis, we did not want to restrict our study to a small, fixed set of LLMs. Instead, we began by selecting a diverse pool of recent and high-performing LLMs. Candidate models were selected based on availability, recency, reported performance on public benchmarks, and suitability for instruction-following and reasoning-oriented tasks. This initial pool included both proprietary and open-weight models accessed through the Gemini API and Google Cloud Vertex AI: Gemini 3.1 Pro, Gemini 3 Flash, Gemini 2.5 Pro, Gemini 2.5 Flash, Gemma 4 26B-A4B, GPT-OSS-120B, GPT-OSS-20B, DeepSeek R1, GLM-5, and Qwen3.5 35B-A3B. 

The pilot study considered all candidate models. After the pilot study, we selected six model configurations for full evaluation. The final set includes three proprietary models and three open-weight models: \textit{Proprietary:} Gemini 3.1 Pro~(\texttt{gemini-3.1-pro-preview}), Gemini 3 Flash~(\texttt{gemini-3-flash-preview}), Gemini 2.5 Pro~(\texttt{gemini-2.5-pro}). \textit{Open-weight:} DeepSeek R1~(\texttt{deepseek-ai/deepseek-r1-0528-maas}), GPT-OSS-120B~(\texttt{openai/gpt-oss-120b-maas}), GPT-OSS-20B~(\texttt{openai/gpt-oss-20b-maas}).

These six models were chosen to satisfy three criteria. First, each model is capable of producing reasoning-style outputs and model-generated rationales. Second, the set spans three independent model providers (Google, OpenAI, and DeepSeek-AI) to mitigate single-vendor or single-family bias in our findings. Third, we include a balanced mix of proprietary and open-weight models to enable more reproducible experimentation. This design follows emerging guidance for empirical studies involving LLMs in software engineering, which emphasizes transparency about prompts and system design, reporting interaction traces, using appropriate baselines and metrics, and validating LLM outputs against human judgment~\cite{baltes2025guidelines}. All models are evaluated under identical experimental conditions: the same set of issue reports, the same prompt variants, and the same issue representation. We use a low fixed temperature of \texttt{0.2} to reduce variability while still allowing models to produce natural-language rationales, with \texttt{top-p = 0.95}. For models exposing \texttt{top-k}, we set it to \texttt{64}. 

Each model is instructed to produce a Markdown response containing the following sections: a step-by-step rationale (\textit{Reasoning}), a three-way \textit{Probability Distribution} over the bug classes, a categorical \textit{Confidence} level (High/Medium/Low), and a single-label \textit{Final Answer}. We do not enforce native structured output such as JSON mode. Instead, we preserve the model's natural explanatory response because the generated rationale is part of the empirical material analyzed in RQ2. The final label, confidence level, and probability distribution are extracted using regular-expression parsing. The full response text is retained for qualitative analysis.

\subsection{Prompt Refinement}
\label{sec:promptdesign}

\noindent\textbf{Baseline Prompt:} We adopt the classification prompt from the replication package accompanying Version~3.0 of the InEx-Bug dataset~\cite{wright2025inexbug} as our baseline (\texttt{v0}). The prompt instructs the model to classify each GitHub issue as \textit{Intrinsic Bug}, \textit{Extrinsic Bug}, or \textit{Not a Bug}, using formal definitions, decision rules, and seven few-shot examples. Before use, we corrected only structural inconsistencies, 
without adding, removing, or semantically altering any definitions, rules, or examples.

\noindent\textbf{Prompt Variants:} We create twelve variants of the baseline prompt to study how domain-specific rules and general prompt-engineering changes affect classification. The first set uses \textit{domain-informed ablations}. We selectively remove rules that may overfit to specific issue patterns, such as upstream-library attribution or temporal-signal heuristics. These rules were selected during a manual review of the baseline prompt. The second set uses \textit{general prompt-engineering variants}, modifying prompt structure, wording, or ordering while preserving domain content, motivated by prior work on prompt sensitivity~\cite{lu2022fantastically,mizrahi2024state,salinas2024butterfly}.

Our variants were constructed along four dimensions:
\begin{itemize}
    \item \textit{Domain-informed ablation}: Remove domain-specific rules to assess their impact on classification. 
    \item \textit{Structural ablation}: Remove structural components (e.g., few-shot examples and probability outputs) to measure their impact on classification.
    \item \textit{Structural sensitivity}: Test whether formatting or ordering changes affect classification performance.
    \item \textit{Linguistic sensitivity}: Test whether tone, register, or persona changes affect classification performance.
\end{itemize}

\noindent\textbf{Pilot Study:} We conduct a pilot study to select prompt variants for full-scale evaluation. The pilot uses a stratified sample of 30 labeled issue reports, consisting of 10 \textit{Intrinsic Bug}, 10 \textit{Extrinsic Bug}, and 10 \textit{Not a Bug} reports, drawn with a fixed random seed for reproducibility. Each of the 13 prompt variants was evaluated on this sample using 24 configurations from 10 proprietary and open-weight models.

Prompt performance is assessed using overall accuracy and \textit{Extrinsic Bug} recall, aggregated across all 24 model configurations. We emphasize \textit{Extrinsic Bug} recall because extrinsic bugs are the most domain-sensitive class. 
A prompt that improves overall accuracy while failing to identify extrinsic bugs would therefore not adequately address the task. All three authors jointly reviewed the accuracy and \textit{Extrinsic Bug} recall results across the 13 variants and reached consensus on the three variants that performed most consistently well across models and were worth investigating further: the \textit{baseline} prompt (\texttt{v0}), the \textit{no-upstream-library-rule} variant (\texttt{v10}), and the \textit{maintainer-voice} variant (\texttt{v11}), referred to as \texttt{v0}, \texttt{v1}, and \texttt{v2} in the full evaluation study.

\subsection{RQ1: Prompt Evaluation}
\label{sec:rq1method}
We evaluate the three selected prompt variants on the full dataset of 428 issue reports using the six selected LLMs. This produces 18 (model~$\times$~prompt) configurations in total.

\noindent\textbf{Mitigating LLM Non-Determinism:} Since LLM outputs can vary across repeated runs~\cite{ouyang2025tosem, atil2025arxiv}, we executed each model~$\times$~prompt configuration three independent times at \texttt{temperature=0.2}, yielding 54 model~$\times$~prompt~$\times$~run configurations in total. We assess run-to-run stability using: (i) per-instance label agreement reported as the fraction of the 428 issues for which all three runs produced identical predictions, (ii) Fleiss'~$\kappa$ with bootstrap 95\% confidence intervals~\cite{fleiss1971} treating the three runs as raters, (iii) mean, standard deviation, and min--max range of weighted F1, macro F1, and per-class recall across all runs, and (iv) Cochran's~Q tests on the per-instance binary correctness vector, where an instance was coded as correct if \texttt{ground\_truth == predicted\_label} and incorrect otherwise, to detect statistically significant differences in correctness rate across runs. Benjamini-Hochberg correction controlled the false discovery rate~\cite{benjamini1995}.

\noindent\textbf{Aggregating Across Runs:} We report performance in two ways: (i) per-run mean and standard deviation and (ii) majority-vote (aggregation inspired by self-consistency~\cite{wang2023selfconsistencyimproveschainthought}), where for each issue, the final label is the one predicted by at least two of three runs. In rare cases where all three runs produce a different label, the tie is broken by selecting the label whose run reported the highest predicted class probability. If probabilities are equal, we fall back to lexicographic order to ensure deterministic and reproducible results. 

\noindent\textbf{Evaluation Metrics:} We evaluate classification performance using accuracy, weighted F1, macro F1, and per-class precision, recall, and F1 for each of the three classes. To test prompt differences, we apply McNemar's test to all three pairwise prompt comparisons within each model using majority-vote predictions. 
McNemar's test is appropriate here because the predictions are paired at the issue level and the outcome is binary (correct/incorrect). Benjamini--Hochberg correction controls the false discovery rate at $\alpha=0.05$~\cite{benjamini1995}.

\subsection{RQ2: Qualitative Analysis of Recurring Failure Patterns}
\label{sec:rq2method}

To answer RQ2, we analyze the model outputs from RQ1. Each output includes the issue report, ground-truth label, model prediction, rationale, probability distribution, confidence level, prompt variant, model identifier, and run identifier. Since the full corpus contains 23,112 outputs across three repeated runs, we reduce it to one representative response for each (issue~$\times$~model~$\times$prompt). We retain the earliest response matching the majority prediction across the three runs; when all runs disagree, we use run~1 for deterministic selection. This produces 7,704 records.

We analyze this reduced corpus in two complementary ways. First, we compute quantitative summaries of error distributions across models, prompts, classes, and issue reports to identify which classes and decision boundaries are most error-prone. We also categorize issues by classification difficulty to guide qualitative sampling. An issue is labeled \texttt{model\_hard} if at most two of the six models classify it correctly, \texttt{model\_easy} if at least four models classify it correctly, and \texttt{model\_mixed} if exactly three models classify it correctly. This yields 93 hard, 316 easy, and 19 mixed issues. 

\begin{table}[h]
\centering
\caption{Rationale-inspection template used for RQ2.}
\label{tab:rationale_inspection_template}
\scriptsize
\setlength{\tabcolsep}{3pt}
\renewcommand{\arraystretch}{0.92}
\begin{adjustbox}{max width=\columnwidth}
\begin{tabularx}{\columnwidth}{
  >{\raggedright\arraybackslash}p{0.12\columnwidth}
  >{\raggedright\arraybackslash}p{0.32\columnwidth}
  >{\raggedright\arraybackslash}X
}
\toprule
\textbf{ID} & \textbf{Rationale Feature} & \textbf{Options} \\
\midrule

L1-Q1 & Maintainer disagreement & Conflict acknowledged and reasoned through; Conflict present but not explicitly reasoned; No maintainer conflict present. \\

L1-Q2 & Rule grounding & Grounded; Asserted; Not applicable, No rule invoked. \\

L2-Q1 & Rationale and response alignment & Aligned; Compressed; Reversed; Partial self-correction. \\ 

L2-Q2 & Alternative-label consideration & Both alternatives; One alternative; None. \\ 

L2-Q3 & Contradictory evidence handling & Addressed; Dismissed without supporting reasoning; Ignored entirely; No contradicting evidence present. \\

L2-Q4 & Correct-label abandonment & Yes; No. \\

L2-Q5 & Direction of reasoning & Evidence-first; Conclusion-first; Recursive; Reversed. \\

\bottomrule
\end{tabularx}
\end{adjustbox}
\end{table}

\noindent\textbf{Qualitative Inspection:} We used an iterative manual qualitative inspection process to develop a structured template for identifying recurring patterns in model-generated rationales. The purpose of this process was not to conduct open-ended manual coding of the entire response corpus or a sample of it. Instead, our goal was to inspect a focused set of model-generated outputs, identify recurring rationale-level and prompt-interaction patterns, and translate these observations into a structured template that could be applied consistently at scale.

The inspection process proceeded in two rounds. In the first round, two authors inspected a randomly selected sample of 10 issue reports, consisting of five \texttt{model\_easy} and five \texttt{model\_hard} cases. The sample included issue reports from all three ground-truth categories. 
For each sampled issue, the authors individually inspected one representative model output selected from the reduced corpus, together with the issue evidence, the predicted label, and the ground-truth label. They noted recurring behaviors in how the model interpreted the issue, considered alternative labels, used or ignored contextual cues, and justified its prediction. After the independent inspection, the authors met to compare observations, discuss emerging patterns, resolve disagreements through discussion, and refine the candidate template questions. In the second round, the authors repeated this process on a different randomly selected sample of 10 issue reports, again consisting of five \texttt{model\_easy} and five \texttt{model\_hard} cases. Based on the observations and discussions from both rounds, the two authors formalized a preliminary inspection template consisting of structured questions and predefined answer choices. They then applied the preliminary template to the 20 inspected cases from Rounds 1 and 2 to assess whether the questions and answer choices captured the behaviors observed during manual inspection. The third author subsequently reviewed the template and helped refine ambiguous questions and response options. Across these iterations, the authors consolidated overlapping observations, clarified the wording of template questions, and revised the answer choices to better capture recurring patterns in the model-generated rationales. This process resulted in a finalized rationale-inspection template designed to identify recurring patterns associated with misclassifications and to guide the domain-informed pipeline design for RQ3. Table~\ref{tab:rationale_inspection_template} summarizes the template questions and predefined answer options.

\noindent\textbf{LLM-Assisted Template Application:} 
The finalized rationale-inspection template was converted into an LLM-assisted inspection prompt. We used Gemini~2.5~Pro to apply the finalized template because it was the best-performing model in the RQ1 classification evaluation. Gemini~2.5~Pro was configured with a maximum thinking budget of 32,768 tokens. The model was instructed to produce one structured CSV row per response record, containing the seven template values in a fixed column order. We applied the inspection prompt to the full set of 7,704 representative responses. To assess the consistency of the LLM-assisted template application, we compared its outputs against human-coded labels on two disjoint ten-record samples corresponding to the two human-inspection rounds. The comparison showed cell-level agreement above 90\%.

\subsection{RQ3: Domain-Informed Multi-Stage Pipeline}
\label{sec:rq3method}
To answer RQ3, we build a domain-informed multi-stage classification pipeline using the strongest RQ1 prompt-only configuration, the \textit{no-upstream-library-rule} variant (\texttt{v1}), as the baseline. Guided by the RQ2 failure analysis, the pipeline separates bug validity from bug origin and adds targeted checks for validity errors.

\noindent\textbf{Task Decomposition:} The prompt-only baseline treats classification as a single three-way decision. In contrast, our pipeline decomposes the task into two sequential binary decisions. The first stage determines bug validity, while the second stage determines bug origin only for issues classified as valid bugs.

\begin{enumerate}
    \item \textbf{Stage 1: Bug Validity} (\textit{Bug} vs \textit{Not a Bug}) determines whether the report describes a genuine software defect requiring a change to the package's shipped code, configuration, or constraints, regardless of where the defect originates.
    \item \textbf{Stage 2: Bug Origin} (\textit{Intrinsic Bug} vs \textit{Extrinsic Bug}) applies only to reports classified as bugs. This stage determines whether the defect originates from the package's own logic or from an external dependency, environment, platform, or upstream change.
\end{enumerate}

The two-stage pipeline first asks whether a defect exists and then, only if a defect exists, asks where the defect originated. The Stage~1 prompt treats both internally and externally caused defects as \textit{Bug}, does not assign the origin label, and instead defers the origin analysis to Stage~2. The Stage~2 prompt then guides the model to reason about where the defect originated and not where the fix may be applied. By decomposing the three-way classification task into two focused binary decisions, the pipeline avoids forcing a single prompt to arbitrate bug validity and bug origin at the same time. 

\noindent\textbf{Domain-Informed Stage 1:}
The Stage~1 defect prompt is adopted from the \texttt{v1} prompt selected in RQ1. We add two rules derived from the RQ2 analysis of how \textit{Extrinsic Bug} reports are incorrectly dismissed. RQ2 showed that many \textit{Extrinsic Bug} errors occur when a model recognizes an externally triggered failure but revises away from the bug label because the focal project is not directly responsible for the underlying cause. To address this failure mode, the Stage~1 prompt emphasizes that an issue can describe a genuine bug even when the root cause lies outside the focal repository, and that bug validity should be decided before assigning responsibility for the origin of the defect. 

\noindent\textbf{Independent Re-check Stage (\texttt{Recover}):} Between Stage~1 and Stage~2, we insert a re-check stage, called \texttt{Recover}.  When Stage~1 predicts \textit{Not a Bug}, \texttt{Recover} examines whether the issue nevertheless describes a genuine software defect that should instead be classified as a bug. \texttt{Recover} receives the Stage~1 verdict and rationale as a ``prior pipeline assessment'', but is instructed to treat that assessment as a prior opinion that may be wrong rather than as a decision to ratify. This stage is conservative by design: it retains the Stage~1 verdict unless the issue contains clear evidence that the verdict is wrong. The re-check is performed by an independent judge model. In the preliminary pipeline iterations, using the same model for both Stage~1 and the re-check often led the model to re-derive and re-endorse its earlier decision. \texttt{Recover} is performed by \texttt{gemini-3.1-pro-preview} for all pipelines except the one whose Stage~1 also uses \texttt{gemini-3.1-pro-preview}. There, to keep the judge independent, we substitute \texttt{gemini-2.5-pro}.

\noindent\textbf{Context Chaining between Stages:} Each stage is executed as an independent LLM inference and therefore has no conversational memory. To preserve intermediate decisions, downstream stages receive the verdict and rationale generated by preceding stages. The \texttt{Recover} stage receives the Stage~1 prediction and rationale. Stage~2 receives the bug-validity verdict and the preceding rationale, and then assumes that bug validity has already been established. It then determines whether the defect originates from within the focal package (intrinsic) or from an external source (extrinsic). By carrying the accumulated reasoning forward in this pipeline, it allows for the avoidance of collapsing stages within a single run and allows for early mistakes to be caught later in the pipeline.  

\noindent\textbf{Evaluation:} 
For comparability with RQ1, we run the proposed pipeline end-to-end on the same 428-report dataset used in the prompt-only evaluation. Each issue receives a final three-way label: \textit{Intrinsic Bug}, \textit{Extrinsic Bug}, or \textit{Not a Bug}. We then evaluate the proposed pipeline on an independent validation set of 30 manually annotated npm issue reports, with 10 reports per class. This validation set was not used during prompt selection, rationale-template development, failure analysis, or pipeline construction. The validation reports were manually annotated by the authors using the same three-way schema as InEx-Bug, with disagreements resolved through discussion. Lastly, we perform component ablations to isolate the contribution of each pipeline element. For the \texttt{Recover} stage, we record recovery flips and whether each flip improves or harms the final classification. Classifications that did not return a response or did not contain a parsable classification label were treated as invalid predictions and counted as incorrect.

\section{Study Results}
\label{sec:studyresults}

Before addressing the research questions, we summarize the pilot study used to select prompt variants for full-scale evaluation. In the pilot study, accuracy ranged from 63.9\% (\texttt{v3}) to 68.2\% (\texttt{v10}) across the prompt variants. \textit{Extrinsic Bug} recall varied from 17.1\% (\texttt{v5}) to 35.4\% (\texttt{v10}). Prompt \texttt{v10} achieved the highest accuracy (+1.8 pp over \texttt{v0}) and the largest gain in \textit{Extrinsic Bug} recall (+12.5 pp). Although \texttt{v12}, which combines \texttt{v10} and \texttt{v11}, achieved similar gains, it did not outperform \texttt{v10}. We therefore selected the baseline prompt (\texttt{v0}), the no-upstream-library-rule variant (\texttt{v10}), and the maintainer-voice variant (\texttt{v11}) for full-scale evaluation. For readability, we refer to these three prompts in the full evaluation as \texttt{v0}, \texttt{v1}, and \texttt{v2}, respectively.

\subsection{RQ1: Prompt-Based Issue Classification}
\label{sec:rq1results}

\begin{table}[t]
\centering
\caption{RQ1 classification performance. Macro F1 is reported on majority-vote predictions} 
\footnotesize
\label{tab:rq1_results}
\begin{tabular}{lcccccc}
\toprule
& \multicolumn{2}{c}{\textbf{v0}}
& \multicolumn{2}{c}{\textbf{v1}}
& \multicolumn{2}{c}{\textbf{v2}} \\
\cmidrule(lr){2-3}\cmidrule(lr){4-5}\cmidrule(lr){6-7}
\textbf{Model} & F1 & Acc & F1 & Acc & F1 & Acc \\
\midrule
DeepSeek-R1 & 0.689 & 0.729 & \textbf{0.735}$^\dagger$ & \textbf{0.752} & 0.704 & 0.738 \\
Gemini-2.5-Pro & 0.722 & 0.755 & \textbf{0.801}$^\dagger$ & \textbf{0.815} & 0.743 & 0.776 \\
Gemini-3-Flash & 0.649 & 0.717 & \textbf{0.769}$^\dagger$ & \textbf{0.787} & 0.669 & 0.727 \\
Gemini-3.1-Pro & 0.626 & 0.706 & \textbf{0.733}$^\dagger$ & \textbf{0.762} & 0.631 & 0.708 \\
GPT-OSS-120B & 0.765 & 0.783 & \textbf{0.779} & \textbf{0.792} & 0.765 & 0.783 \\
GPT-OSS-20B & 0.718 & 0.748 & 0.753 & 0.764 & \textbf{0.755} & \textbf{0.773} \\
\bottomrule
\end{tabular}
\begin{tablenotes}
\footnotesize
\item$^\dagger$ Significant improvement over \texttt{v0} under McNemar's test with BH-FDR correction ($\alpha=0.05$).
\end{tablenotes}
\end{table}

\noindent\textbf{Run Stability Analysis:}  Across all 18 (model~$\times$~prompt) configurations, at least two of the three runs agreed on the predicted label for 98.8\%--100\% of issues (mean: 99.7\%), confirming that majority vote labels were rarely determined by a single run. Full agreement ranged from 81.5\% to 97.2\% (mean: 89.5\%). Treating runs as raters, Fleiss'~$\kappa$ ranged from 0.804 to 0.963 across all 18 configurations (mean: 0.883), with all configurations exceeding the conventional threshold for almost-perfect agreement $\kappa = 0.80$~\cite{landis1977}. Cochran's~Q tests on the per-instance binary correctness detected no significant differences across the runs after global BH-FDR correction. 
Class-conditional stability is the fraction of issues per ground-truth class for which all three runs agreed. \textit{Extrinsic Bug} instances were least stable (mean: 80.1\%), compared with \textit{Intrinsic Bug} (94.1\%) and \textit{Not a Bug} (91.9\%).

\noindent\textbf{Classification Results:} Given the high inter-run agreement across all configurations ($\kappa \geq 0.804$), we report RQ1 classification results using majority-vote predictions as point estimates, together with per-run mean~$\pm$~SD. Table~\ref{tab:rq1_results} shows macro F1 and accuracy for all 18 model~$\times$~prompt configurations. Macro F1 on majority-vote (MV) predictions ranged from 0.626 (Gemini-3.1-Pro~$\times$~v0) to 0.801 (Gemini-2.5-Pro~$\times$~v1), with per-run mean~$\pm$~SD confirming low sensitivity to any individual run across all configurations. The best-performing model was Gemini-2.5-Pro~$\times$~v1 (macro F1 = 0.801, weighted F1 = 0.815, accuracy = 81.5\%), followed by GPT-OSS-120B~$\times$~v1 (macro F1 = 0.779) and Gemini-3-Flash~$\times$~v1 (macro F1 = 0.769).

Prompt v1 achieved the highest macro F1 for five of six models, with gains over v0 for all six models ranging from +1.4 pp (GPT-OSS-120B) to +11.9 pp (Gemini-3-Flash). The only model for which v1 was not the top-performing prompt was GPT-OSS-20B, where v2 marginally but non-significantly outperformed v1 by 0.2 pp (macro F1 = 0.755 vs.\ 0.753). For four of the six models, Gemini-2.5-Pro, Gemini-3-Flash, Gemini-3.1-Pro, and DeepSeek-R1, the v0-to-v1 gains were statistically significant under McNemar's test with BH-FDR correction ($p_{\text{FDR}} \leq 0.022$).

\textit{Extrinsic Bug} recall showed the largest and most consistent gains from \texttt{v0} to \texttt{v1} across all models, ranging from +9.3 pp (GPT-OSS-120B) to +35.5 pp (Gemini-3-Flash). Under \texttt{v0}, \textit{Extrinsic Bug} recall was particularly low for the Gemini-3.1-Pro (0.215) and Gemini-3-Flash (0.262) models. 

\subsection{RQ2: Qualitative Analysis of Recurring Failure Patterns}
\label{sec:rq2results}
Across all 7,704 coded records, LLMs produced correct classifications in 5,829 cases (75.7\%) and incorrect ones in 1,875 (24.3\%). Error rates differed substantially across ground-truth categories: \textit{Extrinsic Bug} issues were misclassified in 975 of 1,926 records (50.6\%), compared with 235 of 1,926 (12.2\%) for \textit{Intrinsic Bug} and 665 of 3,852 (17.3\%) for \textit{Not a Bug}. Table~\ref{tab:misclass} summarizes the six misclassification directions by volume. The two largest error directions are \textit{Extrinsic Bug} predicted as \textit{Not a Bug} ($n=609$) and \textit{Not a Bug} predicted as \textit{Intrinsic Bug} ($n=487$), together accounting for around 58\% of all misclassifications.

\begin{table}[t]
  \centering
  \caption{Misclassification counts by direction (all 7,704 records).}
  \label{tab:misclass}
  \begin{tabular}{lrrr}
    \toprule
    Direction & $n_{\text{wrong}}$ & $n_{\text{correct GT}}$ & \% of all wrong \\
    \midrule
    \textbf{Extrinsic $\to$ Not a Bug}   & \textbf{609} & \textbf{951}   & \textbf{32.5}\% \\
    Not a Bug $\to$ Intrinsic   & 487 & 3,187 & 26.0\% \\
    Extrinsic $\to$ Intrinsic   & 366 & 951   & 19.5\% \\
    Intrinsic $\to$ Not a Bug   & 185 & 1,691 &  9.9\% \\
    Not a Bug $\to$ Extrinsic   & 178 & 3,187 &  9.5\% \\
    Intrinsic $\to$ Extrinsic   &  50 & 1,691 &  2.7\% \\
    \bottomrule
  \end{tabular}
\end{table}

\begin{figure}[t]
  \centering
  \includegraphics[width=\columnwidth]{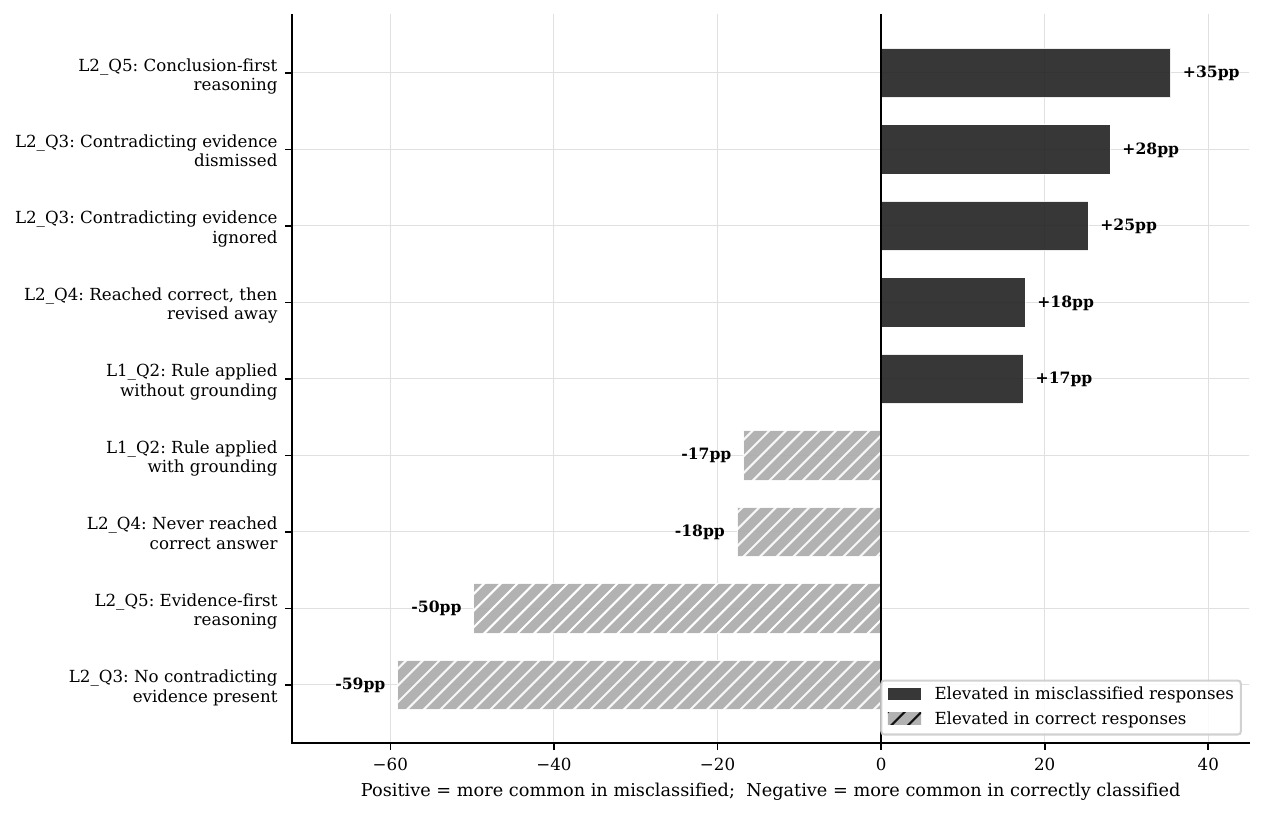}
  \caption{Discrimination of coding questions between correct and incorrect predictions.}
  \label{fig:diverging}
\end{figure}

\medskip
\noindent\textbf{Cross-Class Failure Patterns.}
Two coding questions from the rationale-inspection template distinguish correct predictions from the incorrect ones with consistency across all three ground-truth classes: L2-Q3 (Contradictory evidence handling) and L2-Q5 (Direction of reasoning). Figure~\ref{fig:diverging} visualizes the full discriminating profile.

For L2-Q3, 63.4\% of correct responses are coded as having \emph{no contradicting evidence present} 
compared with only 4.3\% of misclassified responses ($\Delta = -59$\,pp). In misclassified responses, contradictory evidence was much more often ignored or dismissed. Specifically, 31.0\% ignore it entirely ($\Delta = +25$\,pp vs.\ 5.7\% in correct responses), and 29.1\% acknowledged but dismissed it without supporting reasoning ($\Delta = +28$\,pp vs.\ 1.1\%). This pattern appeared across all three ground-truth classes: \textit{Intrinsic Bug} ($\Delta = -66$\,pp), \textit{Not a Bug} ($\Delta = -59$\,pp), \textit{Extrinsic Bug} ($\Delta =-47$\,pp).

For L2-Q5, evidence-first reasoning appears in 91.4\% of correct responses but only 41.6\% of misclassified ones ($\Delta = -49$\,pp). Conversely, conclusion-first reasoning appears in only 5.4\% of correct responses but 40.8\% of misclassified responses ($\Delta = +35$\,pp). This pattern is consistent across all three categories: \textit{Extrinsic Bug} ($\Delta = +33$\,pp), \textit{Intrinsic Bug} ($\Delta = +42$\,pp), and \textit{Not a Bug} ($\Delta = +35$\,pp). 

\noindent\textbf{\textit{Extrinsic Bug} Exhibits a Distinct Rationale Profile.}
\textit{Extrinsic Bug} issues exhibit a different rationale profile from the other two classes. Conclusion-first (L2-Q5) reasoning appears in 22.4\% \textit{Extrinsic Bug} responses, compared with 12.0\% for \textit{Intrinsic Bug} and 10.8\% for \textit{Not a Bug}. Models also revise away (L2-Q4) from the correct label more often for \textit{Extrinsic Bug} cases (10.3\% vs.\ 2.3\% for both other classes). Finally, contradicting evidence (L2-Q3) is more common in \textit{Extrinsic Bug} records (71.9\%) than in \textit{Intrinsic Bug} (40.2\%) or \textit{Not a Bug} (45.9\%), leaving fewer cases with no contradicting evidence (28.1\% vs.\ 59.8\% and 54.1\%).

\noindent\textbf{Observed Patterns by Direction of Misclassifications.}
When \textit{Extrinsic Bug} issues are misclassified as \textit{Intrinsic Bug} ($n=366$), the dominant profile combines conclusion-first reasoning with asserted rule application. Conclusion-first reasoning (L2-Q5) appears in 51.6\% of these wrong responses versus 5.3\% of correct \textit{Extrinsic Bug} responses ($\Delta = +46$\,pp). L1-Q2 is coded \emph{rule asserted without grounding} in 35.5\% of wrong responses versus 1.7\% of correct ones ($\Delta = +33$\,pp). Contradicting evidence is dismissed in 35.2\% of cases ($\Delta = +33$\,pp
vs.\ 1.5\%). When \textit{Extrinsic Bug} issues are misclassified as \textit{Not a Bug} ($n=609$), 27.1\% of responses first reach the correct label before revising away from it (L2-Q4), compared with 0.1\% of correct \textit{Extrinsic Bug} responses ($\Delta = +27$\,pp). Conclusion-first reasoning ($\Delta = +26$\,pp) and dismissed contradicting evidence ($\Delta = +23$\,pp) are also elevated. 

When \textit{Not a Bug} cases are predicted as \textit{Intrinsic Bug} ($n=487$), \emph{one alternative considered (L2-Q2)} appears in 33.5\% of wrong responses versus 9.0\% of correct \textit{Not a Bug} responses ($\Delta = +24$\,pp). In contrast, \emph{no alternatives considered} drops from 41.1\% (correct) to 18.1\% (wrong) ($\Delta = -23$\,pp).

\subsection{RQ3: Domain-Informed Multi-Stage Pipeline}
\label{sec:rq3results}

Table \ref{tab:ablation_all_models_compact} reports the proposed pipeline results across all models. The proposed pipeline improves macro F1 over the prompt-only baseline for five of the six models, with gains ranging from +0.8 pp (DeepSeek-R1) to +4.8 pp (Gemini-3-Flash). The \textit{Extrinsic Bug} recall also improves for five of the six models, most sharply for Gemini-3.1-Pro (0.505$\to$0.738) and Gemini-3-Flash (0.617$\to$0.757). The strongest prompt-only configuration (baseline) was Gemini-2.5-Pro (accuracy 0.815, \textit{Extrinsic Bug} recall 0.682). The strongest pipeline configuration, Gemini-3-Flash, reaches 0.827 accuracy, a +1.2 pp improvement, while simultaneously raising \textit{Extrinsic Bug} recall to 0.757, +7.5 pp above the best-accuracy baseline.

Each stage is independently reliable. Table \ref{tab:stage_breakdown} separates the two decisions. Stage~2 (\textit{Intrinsic Bug} vs. \textit{Extrinsic Bug} origin) is the more accurate stage, reaching 0.864–0.907 across all six models. Stage~1 (\textit{Bug} vs. \textit{Not a Bug} validity) reaches 0.785–0.871, with high \textit{Bug} recall (0.813–0.930) but lower \textit{Not a Bug} recall (0.720–0.836). The \texttt{Recover} pass produces a net-positive number of beneficial flips on five of six models, harming only DeepSeek-R1.

\begin{table*}[t]
\centering
\caption{Results across models and pipeline configurations. Each cell reports Macro-F1 / Accuracy / Extrinsic Recall.}
\label{tab:ablation_all_models_compact}
\scriptsize
\setlength{\tabcolsep}{3pt}
\renewcommand{\arraystretch}{1.15}

\begin{tabularx}{\textwidth}{lXXXXXX}
\toprule
\textbf{Configuration}
& \textbf{Gemini-2.5-Pro}
& \textbf{Gemini-3.1-Pro}
& \textbf{Gemini-3-Flash}
& \textbf{GPT-OSS-120B}
& \textbf{GPT-OSS-20B}
& \textbf{DeepSeek-R1} \\
\midrule

Best prompt-only v1 (baseline)
& 0.801 / 0.815 / 0.682
& 0.733 / 0.762 / 0.505
& 0.769 / 0.787 / 0.617
& 0.779 / 0.792 / 0.692
& 0.753 / 0.764 / 0.682
& 0.735 / 0.752 / 0.607 \\

A: Stage 1 (v1) + Stage 2 (v1) 
& 0.776 / 0.794 / 0.692
& 0.730 / 0.766 / 0.458
& 0.759 / 0.783 / 0.589
& 0.778 / 0.792 / 0.682
& 0.787 / 0.801 / 0.682
& 0.743 / 0.738 / 0.589 \\

B: A + context chaining
& 0.789 / 0.804 / 0.692
& 0.719 / 0.757 / 0.449
& 0.776 / 0.797 / 0.598
& 0.787 / 0.799 / 0.710
& 0.765 / 0.783 / 0.607
& 0.710 / 0.713 / 0.505 \\

C: B +  \texttt{Recover}
& 0.799 / 0.813 / 0.701
& 0.752 / 0.762 / 0.692
& 0.798 / 0.815 / 0.664
& 0.788 / 0.801 / 0.720
& 0.787 / 0.799 / 0.673
& 0.734 / 0.717 / 0.579 \\

Proposed pipeline: \\
C with domain-informed Stage 1 
& 0.801 / 0.815 / 0.738
& 0.765 / 0.771 / 0.738
& \textbf{0.817} / \textbf{0.827} / \textbf{0.757}
& 0.802 / 0.811 / 0.757
& 0.769 / 0.783 / 0.682
& 0.743 / 0.734 / 0.654 \\

\bottomrule
\end{tabularx}
\end{table*}

\begin{table}[t]
\centering
\caption{Stage-level breakdown of the proposed pipeline. 
}
\footnotesize
\label{tab:stage_breakdown}
\setlength{\tabcolsep}{3pt}
\begin{tabular}{lcccccc}
\toprule
& \multicolumn{3}{c}{\textbf{Stage 1}}
& \multicolumn{1}{c}{\textbf{Stage 2}}
& \multicolumn{2}{c}{\textbf{Recovery}} \\
\cmidrule(lr){2-4}\cmidrule(lr){5-5}\cmidrule(lr){6-7}
\textbf{Model} & Acc & Bug R & NAB R & Acc & Flips \\
\midrule
\textbf{Gemini-2.5-Pro} & \textbf{0.871} & 0.916 & 0.827 & 0.878 & 4 (4/0/+4)\\
Gemini-3.1-Pro & 0.825 & 0.930 & 0.720 & 0.884 & 65 (34/31/+3) \\
\textbf{Gemini-3-Flash} & 0.869 & 0.902 & 0.836 & \textbf{0.907} & 6 (4/2/+2)  \\
GPT-OSS-120B & 0.855 & 0.925 & 0.785 & 0.904 & 6 (5/1/+4)  \\
GPT-OSS-20B & 0.843 & 0.893 & 0.794 & 0.864 & 8 (5/3/+2)  \\
DeepSeek-R1 & 0.785 & 0.813 & 0.757 & 0.874 & 7 (3/4/$-$1)  \\
\bottomrule
\end{tabular}
\begin{tablenotes}
\footnotesize
\item Bug R = \textit{Bug} recall; NAB R = \textit{Not a Bug} recall. Recovery flips are reported as total flips with good/bad/net in parentheses.
\end{tablenotes}
\end{table}

\noindent\textbf{Validation Set:}  On the independent 30-issue validation set, Macro F1 ranges 0.829–0.935 and accuracy 0.800–0.933, with GPT-OSS-120B and GPT-OSS-20B achieving the highest macro F1 (0.935 and 0.933). \textit{Not a Bug} recall stays high (0.900–1.000), while \textit{Extrinsic Bug} recall (0.700–1.000) remains the most variable, consistent with the test-set results.

 
\section{Discussion and Implications}
\noindent\textbf{Prompt-only refinement plateaus, and added domain rules can negatively impact performance.} 
It is worth noting that the strongest prompt variant was not the one that encoded the most domain guidance. In fact, \texttt{v1} was a \textit{domain-informed ablation} that removed the upstream-library-attribution rule. It achieved the highest macro F1 for five of six models and improved over the baseline \texttt{v0} for all six models. 
Our results align with prior evidence that, for LLMs, more context is not necessarily better. Prompt sensitivity shows that small, semantically minor edits to a prompt can swing the classification outcomes~\cite{lu2022fantastically,mizrahi2024state,salinas2024butterfly}. Prior work shows that added context can distract LLMs and reduce performance, whether through irrelevant details, excessive examples, or agent ``cursor rules''~\cite{shi2023largelanguagemodelseasily,oskooei2025manyshotpromptingfailsempirical,JiangNam2026CursorRules}. Our results show that even valid expert-authored domain rules can have this effect when encoded directly in prompts.

\noindent\faHandORight~\underline{Implication 1:} For domain-sensitive tasks, prompt-only refinement can face a diminishing-returns ceiling, and adding more domain rules or context is not necessarily beneficial. Domain guidance should therefore be evaluated empirically rather than assumed to improve performance.

\noindent\textbf{Rationales are useful diagnostic artifacts, but not explanations of internal reasoning.} Our RQ2 analysis used model-generated rationales to identify recurring patterns in incorrect predictions. 
Correct predictions were much more likely to be \textit{evidence-first}, whereas incorrect predictions frequently anchored on a conclusion. Similarly, misclassified responses often ignored or dismissed evidence that conflicted with the final label. These findings should not be interpreted as proof of the model's internal reasoning process. Prior work has shown that chain-of-thought and generated explanations can be unfaithful to the actual decision process~\cite{turpin2023language}. We therefore treat rationales as observable artifacts: they show how the model presents and justifies its interpretation, not necessarily how the model internally arrived at the label. Even under this cautious interpretation, rationales are valuable. They expose recurring failure signatures that aggregate metrics cannot show. 
\noindent\faHandORight~\underline{Implication 2:} Researchers should avoid claiming that a model ``reasoned correctly'' merely because its explanation is plausible. Instead, generated rationales can be treated as diagnostic artifacts that can help identify failure patterns and guide the design of more reliable classification pipelines.

\noindent\textbf{\textit{Extrinsic Bugs} are a distinct challenge, not just a lower-performing class.} Prior bug origin research shows that extrinsic bugs are empirically important and can affect downstream software analytics~\cite{rodriguez2020how,rodriguez2020watch}, while issue-report studies show that bug-labeled reports often contain non-bugs~\cite{herzig2013feature}. 
Our findings connect these concerns by showing that LLMs struggle when bug validity and bug origin must be judged together, even with explicit definitions, examples, and decision rules. The difficulty is not simply that \textit{Extrinsic Bug} is a lower-performing class; it requires the model to recognize that a reported failure can be real even when the focal repository did not introduce the root cause. LLMs also revised away from the correct label more often for \textit{Extrinsic Bug} reports. These results suggest that extrinsic bugs create a specific interpretive burden. This burden appears tied to how such reports are written and discussed, and it persisted despite explicit classification guidance in our evaluation.

\noindent\faHandORight~\underline{Implication 3:} 
Extrinsic bug classification requires richer evidence than labels alone. Future datasets, issue templates, and reporting guidelines should capture dependency, environment, reproduction, maintainer, fix-location, and causal information needed to distinguish external causation from non-bug reports.

\noindent\textbf{For \textit{Extrinsic Bugs}, the main bottleneck is bug validity, not bug origin.} Our pipeline decomposes a complex task into smaller subtasks~\cite{khot2022decomposed} and highlights where the difficulty lies when it comes to classifying \textit{Extrinsic Bugs}. \textit{Extrinsic Bug} was the hardest class to classify, and its dominant error direction is revealing: 32.5\% of all misclassifications were extrinsic bugs mistaken for \textit{Not a Bug}. Once an issue is established as a genuine bug, the origin decision is reliable: Stage 2 achieved up to 90.7\% accuracy on \textit{Intrinsic Bug} vs. \textit{Extrinsic Bug}. The end-to-end performance is instead bounded by Stage 1, where extrinsic bugs are discarded as \textit{Not a Bug} before the origin decision is ever reached. The persistence of this error even under domain-informed prompting suggests the boundary may be genuinely ambiguous, potentially difficult for human annotators as well. This motivates an in-depth study of these cases to identify what evidence distinguishes an externally-caused defect from a non-bug. 

\noindent\faHandORight~\underline{Implication 4:} Datasets, tools, and classification pipelines should treat bug validity and bug origin as separable judgments. Future work should aim to improve extrinsic bug detection by targeting the \textit{Bug} vs. \textit{Not a Bug} boundary.

\noindent\textbf{Comparison with non-LLM classifiers.} Prior automated origin classifiers target a different task, dataset, and label schema. BuggIn~\cite{bhandari2023buggin} performs binary intrinsic-vs-rest classification with classical ML to identify intrinsic bugs (best F1 0.78), and Fan et al.'s TextCNN--KAN~\cite{fan2026AutomaticExtrinsic} performs binary intrinsic-vs-extrinsic classification (accuracy 0.75, F1 0.81 on 1,880 OpenStack samples). LLM-BRC~\cite{du2024llm} and Colavito et al.~\cite{colavito2024impact,colavito2024leveraging,colavito2026issue} classify behavioral or type-level properties rather than bug origin. The closest comparison to our work is at the binary-subtask level: our pipeline reaches 87.1\% accuracy on the validity decision (Stage~1) and 90.7\% on the origin decision (Stage~2), while jointly resolving both judgments that prior classifiers treat in isolation. 

\section{Threats to Validity}

\noindent\textbf{Internal Validity: }
LLM outputs may vary across identical runs, affecting classifications, rationales, and confidence estimates in both the prompt-only evaluation and the downstream analysis. We mitigated this by running each model~$\times$~prompt configuration three times at \texttt{temperature=0.2}, aggregating predictions by majority vote, and measuring stability with agreement, Fleiss'$\kappa$, metric variation, and Cochran's $Q$ tests. These results support using one representative response for RQ2 and single-run pipeline results for RQ3, though residual randomness remains. 

To code 7,704 rationales for RQ2, we applied the finalized template using Gemini 2.5 Pro, the strongest evaluated classifier. An automated coder may misread rationales, and using the model from the same family as some of the coded outputs may bias the coding. We mitigated this by developing the template through an iterative, human-driven inspection process and validating the automated coding against independent human coding on two disjoint ten-record samples, where cell-level agreement exceeded 90\%. However, the validation set remains small. The template also encodes the authors' judgments about which reasoning behaviors are salient, which may shape observed failures. To reduce this, two authors independently inspected samples across two rounds, reconciled disagreements, and a third author refined ambiguous questions and answer options.

\noindent\textbf{External Validity: }
All 428 primary issues come from npm because InEx-Bug is the only public dataset with the required three-way schema. While npm's dependency structure makes extrinsic bugs especially relevant, results may not transfer directly to ecosystems such as Maven or PyPI, industrial codebases, or projects with different triage practices.

\noindent\textbf{Construct Validity: }
Our failure analysis uses model-generated rationales, which may not faithfully reflect internal reasoning~\cite{turpin2023language}. We therefore do not interpret rationales as explanations of why a model decided as it did. Instead, we treat them as observable artifacts of how a model presents its interpretation of the issue. We acknowledge that an inferential gap between stated and actual reasoning remains.
The three classes reflect judgments about bug validity and origin, which can be ambiguous when evidence is incomplete. We treat InEx-Bug annotations as ground truth. Any residual label noise or borderline cases would affect both the performance metrics and the failure analysis.
Finally, because LLMs are pre-trained on large web corpora, some evaluated models may have encountered GitHub issues during pre-training, raising the possibility of training-data leakage. We consider this risk limited, as the threat is not exposure to the issue text, which is the intended model input, but to the ground-truth labels. The manually annotated InEx-Bug dataset was released in 2026 and postdates the pre-training data of most evaluated models. Moreover, if leakage had materially influenced the results, we would expect uniformly high accuracy across all three categories. Instead, we observe a class-dependent error profile. Nevertheless, model training cutoffs and data sources are not fully transparent. We therefore cannot rule out exposure to some issue text and labels during pre-training.

\section{Conclusion}
Overall, our findings show that the value of LLMs for issue-report classification should not be assessed only by asking whether they can assign labels accurately. Prompt-only refinement plateaued at 0.815 accuracy and 0.801 macro F1, but a domain-informed pipeline derived from expert failure analysis improved macro F1 and \textit{Extrinsic Bug} recall for five of the six evaluated models, with its strongest configuration reaching 0.827 accuracy and 0.757 \textit{Extrinsic Bug} recall. Decomposing the task further showed that origin classification is more reliable than validity, indicating that the bug-validity decision may account for much of the remaining error. More importantly, these results show that model failures can reveal domain-specific sources of confusion that, when analyzed by experts, can be translated into a more reliable classification process. This paper therefore demonstrates how expert diagnosis can help identify and address failure modes in LLM-based bug classification.

\bibliographystyle{IEEEtran}
\bibliography{software}

@article{brown2020language,
  title={Language models are few-shot learners},
  author={Brown, Tom and Mann, Benjamin and Ryder, Nick and Subbiah, Melanie and Kaplan, Jared D and Dhariwal, Prafulla and Neelakantan, Arvind and Shyam, Pranav and Sastry, Girish and Askell, Amanda and others},
  journal={Advances in neural information processing systems},
  volume={33},
  pages={1877--1901},
  year={2020}
}

@article{hou2024large,
  title={Large language models for software engineering: A systematic literature review},
  author={Hou, Xinyi and Zhao, Yanjie and Liu, Yue and Yang, Zhou and Wang, Kailong and Li, Li and Luo, Xiapu and Lo, David and Grundy, John and Wang, Haoyu},
  journal={ACM Transactions on Software Engineering and Methodology},
  volume={33},
  number={8},
  pages={1--79},
  year={2024},
  publisher={ACM New York, NY}
}

@inproceedings{antoniol2008bug,
  title={Is it a bug or an enhancement? a text-based approach to classify change requests},
  author={Antoniol, Giuliano and Ayari, Kamel and Di Penta, Massimiliano and Khomh, Foutse and Gu{\'e}h{\'e}neuc, Yann-Ga{\"e}l},
  booktitle={The 2008 conference of the center for advanced studies on collaborative research: meeting of minds},
  pages={304--318},
  year={2008}
}

@article{rodriguez2020watch,
  title={Watch out for extrinsic bugs! a case study of their impact in just-in-time bug prediction models on the openstack project},
  author={Rodr{\'\i}guez-P{\'e}rez, Gema and Nagappan, Meiyappan and Robles, Gregorio},
  journal={IEEE Transactions on Software Engineering},
  volume={48},
  number={4},
  pages={1400--1416},
  year={2020},
  publisher={IEEE}
}

@article{rodriguez2020how,
  title={How bugs are born: a model to identify how bugs are introduced in software components},
  author={Rodr{\'\i}guez-P{\'e}rez, Gema and Robles, Gregorio and Serebrenik, Alexander and Zaidman, Andy and Germ{\'a}n, Daniel M and Gonzalez-Barahona, Jesus M},
  journal={Empirical Software Engineering},
  volume={25},
  number={2},
  pages={1294--1340},
  year={2020},
  publisher={Springer},
  doi={10.1007/s10664-019-09781-y}
}

@inproceedings{herzig2013feature,
  title={It's Not a Bug, It's a Feature: How Misclassification Impacts Bug Prediction},
  author={Herzig, Kim and Just, Sascha and Zeller, Andreas},
  booktitle={Proceedings of the 35th International Conference on Software Engineering},
  pages={392--401},
  year={2013},
  doi={10.1109/ICSE.2013.6606585}
}

@inproceedings{decan2018impact,
  title={On the impact of security vulnerabilities in the npm package dependency network},
  author={Decan, Alexandre and Mens, Tom and Constantinou, Eleni},
  booktitle={Proceedings of the 15th International Conference on Mining Software Repositories},
  pages={181--191},
  year={2018},
  doi={10.1145/3196398.3196401}
}

@inproceedings{kikas2017structure,
  title={Structure and evolution of package dependency networks},
  author={Kikas, Riivo and Gousios, Georgios and Dumas, Marlon and Pfahl, Dietmar},
  booktitle={The 14th International Conference on Mining Software Repositories},
  pages={102--112},
  year={2017},
  organization={IEEE}
}

@inproceedings{karampatsis2020often,
  title={How often do single-statement bugs occur? the manysstubs4j dataset},
  author={Karampatsis, Rafael-Michael and Sutton, Charles},
  booktitle={The 17th international conference on mining software repositories},
  pages={573--577},
  year={2020}
}

@misc{wright2025inexbug,
  title        = {{InEx-Bug}: A Human Annotated Dataset of Intrinsic and Extrinsic Bugs in the npm Ecosystem},
  author       = {Wright, Tanner and Chen, Adams and Rodriguez-Perez, Gema},
  year         = {2025},
  publisher    = {STASER Lab},
  organization = {University of British Columbia},
  address      = {Kelowna, British Columbia, Canada},
  howpublished = {\url{https://github.com/STASER-Lab/InEx-Bug-A-Human-Annotated-Dataset-of-Intrinsic-and-Extrinsic-Bugs-in-the-npm-Ecosystem}},
  note         = {Version 3.0}
}

@inproceedings{msr2026,
  title     = {{InEx-Bug}: A Human Annotated Dataset of Intrinsic and Extrinsic Bugs in the npm Ecosystem},
  author    = {Wright, Tanner and Chen, Adams and Rodr{\'i}guez-P{\'e}rez, Gema},
  booktitle = {Proceedings of the 23rd International Conference on Mining Software Repositories},
  year      = {2026},
  address   = {Rio de Janeiro, Brazil},
  publisher = {ACM},
  note      = {Preprint: https://arxiv.org/abs/2602.13400}
}

@inproceedings{decan2018evolution,
  title={On the Evolution of Technical Lag in the npm Package Dependency Network},
  author={Decan, Alexandre and Mens, Tom and Constantinou, Eleni},
  booktitle={2018 IEEE International Conference on Software Maintenance and Evolution (ICSME)},
  pages={404--414},
  year={2018},
  organization={IEEE}
}

@inproceedings{bhandari2023buggin,
  title={{BuggIn}: Automatic intrinsic bugs classification model using nlp and ml},
  author={Bhandari, Pragya and Rodr{\'\i}guez-P{\'e}rez, Gema},
  booktitle={Proceedings of the 19th International Conference on Predictive Models and Data Analytics in Software Engineering},
  pages={2--11},
  year={2023}
}

@article{cogo2021deprecation,
  title={Deprecation of packages and releases in software ecosystems: A case study on npm},
  author={Cogo, Filipe R and Oliva, Gustavo A and Hassan, Ahmed E},
  journal={Transactions on Software Engineering},
  volume={48},
  number={7},
  pages={2208--2223},
  year={2021},
  publisher={IEEE}
}

@inproceedings{miller2025understanding,
  title={Understanding the Response to Open-Source Dependency Abandonment in the npm Ecosystem},
  author={Miller, Courtney and Jahanshahi, Mahmoud and Mockus, Audris and Vasilescu, Bogdan and Kastner, Christian},
  booktitle={2025 IEEE/ACM 47th International Conference on Software Engineering (ICSE)},
  pages={2355--2367},
  year={2025},
  organization={IEEE}
}

@inproceedings{liu2021automatically,
  title={Automatically detecting and fixing concurrency bugs in go software systems},
  author={Liu, Ziheng and Zhu, Shuofei and Qin, Boqin and Chen, Hao and Song, Linhai},
  booktitle={Proceedings of the 26th ACM International Conference on Architectural Support for Programming Languages and Operating Systems},
  pages={616--629},
  year={2021}
}

@article{du2024llm,
  title={LLM-BRC: A large language model-based bug report classification framework},
  author={Du, Xiaoting and Liu, Zhihao and Li, Chenglong and Ma, Xiangyue and Li, Yingzhuo and Wang, Xinyu},
  journal={Software Quality Journal},
  volume={32},
  number={3},
  pages={985--1005},
  year={2024},
  publisher={Springer}
}

@article{koyuncu2025exploring,
  title={Exploring fine-grained bug report categorization with large language models and prompt engineering: An empirical study},
  author={Koyuncu, Anil},
  journal={ACM Transactions on Software Engineering and Methodology},
  year={2025},
  publisher={ACM New York, NY}
}

@inproceedings{qin2018classifying,
  title={Classifying bug reports into bugs and non-bugs using LSTM},
  author={Qin, Hanmin and Sun, Xin},
  booktitle={Proceedings of the 10th Asia-Pacific Symposium on Internetware},
  pages={1--4},
  year={2018}
}

@article{herbold2020feasibility,
  title={On the feasibility of automated prediction of bug and non-bug issues},
  author={Herbold, Steffen and Trautsch, Alexander and Trautsch, Fabian},
  journal={Empirical Software Engineering},
  volume={25},
  number={6},
  pages={5333--5369},
  year={2020},
  publisher={Springer}
}

@article{hirsch2022using,
  title={Using textual bug reports to predict the fault category of software bugs},
  author={Hirsch, Thomas and Hofer, Birgit},
  journal={Array},
  volume={15},
  pages={100189},
  year={2022},
  publisher={Elsevier}
}

@article{colavito2024impact,
  title={Impact of data quality for automatic issue classification using pre-trained language models},
  author={Colavito, Giuseppe and Lanubile, Filippo and Novielli, Nicole and Quaranta, Luigi},
  journal={Journal of Systems and Software},
  volume={210},
  pages={111838},
  year={2024},
  publisher={Elsevier}
}

@inproceedings{colavito2024leveraging,
  title={Leveraging gpt-like llms to automate issue labeling},
  author={Colavito, Giuseppe and Lanubile, Filippo and Novielli, Nicole and Quaranta, Luigi},
  booktitle={21st International Conference on Mining Software Repositories},
  pages={469--480},
  year={2024}
}

@INPROCEEDINGS{Terdchanakul2017,
  author={Terdchanakul, Pannavat and Hata, Hideaki and Phannachitta, Passakorn and Matsumoto, Kenichi},
  booktitle={2017 IEEE International Conference on Software Maintenance and Evolution}, 
  title={Bug or Not? Bug Report Classification Using N-Gram IDF}, 
  year={2017},
  volume={},
  number={},
  pages={534-538},
  doi={10.1109/ICSME.2017.14}}

@inproceedings{Aktas2023, 
   series={ESEC/FSE ’23},
   title={Issue Report Validation in an Industrial Context},
   DOI={10.1145/3611643.3613887},
   booktitle={Proceedings of the 31st ACM Joint European Software Engineering Conference and Symposium on the Foundations of Software Engineering},
   publisher={ACM},
   author={Aktas, Ethem and Cakmak, Ebru and Inan, Mete and Yilmaz, Cemal},
   year={2023},
   month=nov}

@INPROCEEDINGS{Zhou2014,
  author={Zhou, Yu and Tong, Yanxiang and Gu, Ruihang and Gall, Harald},
  booktitle={2014 IEEE International Conference on Software Maintenance and Evolution}, 
  title={Combining Text Mining and Data Mining for Bug Report Classification}, 
  year={2014},
  volume={},
  number={},
  pages={311-320},
  doi={10.1109/ICSME.2014.53}}

@article{baltes2025guidelines,
  title={Guidelines for empirical studies in software engineering involving large language models},
  author={Baltes, Sebastian and Angermeir, Florian and Arora, Chetan and Bar{\'o}n, Marvin Mu{\~n}oz and Chen, Chunyang and B{\"o}hme, Lukas and Calefato, Fabio and Ernst, Neil and Falessi, Davide and Fitzgerald, Brian and others},
  journal={arXiv preprint arXiv:2508.15503},
  year={2025}
}

@article{colavito2025benchmarking,
  title={Benchmarking large language models for automated labeling: The case of issue report classification},
  author={Colavito, Giuseppe and Lanubile, Filippo and Novielli, Nicole},
  journal={Information and Software Technology},
  volume={184},
  pages={107758},
  year={2025},
  publisher={Elsevier}
}

@article{decan2019empirical,
  title={An empirical comparison of dependency network evolution in seven software packaging ecosystems},
  author={Decan, Alexandre and Mens, Tom and Grosjean, Philippe},
  journal={Empirical Software Engineering},
  volume={24},
  number={1},
  pages={381--416},
  year={2019},
  publisher={Springer}
}

@article{venturini2023depended,
  title={I depended on you and you broke me: An empirical study of manifesting breaking changes in client packages},
  author={Venturini, Daniel and Cogo, Filipe Roseiro and Polato, Ivanilton and Gerosa, Marco A and Wiese, Igor Scaliante},
  journal={ACM Transactions on Software Engineering and Methodology},
  volume={32},
  number={4},
  pages={1--26},
  year={2023},
  publisher={ACM New York, NY, USA}
}

@misc{lu2022fantastically,
      title={Fantastically Ordered Prompts and Where to Find Them: Overcoming Few-Shot Prompt Order Sensitivity}, 
      author={Yao Lu and Max Bartolo and Alastair Moore and Sebastian Riedel and Pontus Stenetorp},
      year={2022},
      eprint={2104.08786},
      archivePrefix={arXiv},
      primaryClass={cs.CL},
      url={https://arxiv.org/abs/2104.08786}, 
}

@misc{mizrahi2024state,
      title={State of What Art? A Call for Multi-Prompt LLM Evaluation}, 
      author={Moran Mizrahi and Guy Kaplan and Dan Malkin and Rotem Dror and Dafna Shahaf and Gabriel Stanovsky},
      year={2024},
      eprint={2401.00595},
      archivePrefix={arXiv},
      primaryClass={cs.CL},
      url={https://arxiv.org/abs/2401.00595}, 
}

@inproceedings{salinas2024butterfly,
  title={The butterfly effect of altering prompts: How small changes and jailbreaks affect large language model performance},
  author={Salinas, Abel and Morstatter, Fred},
  booktitle={Findings of the Association for Computational Linguistics: ACL 2024},
  pages={4629--4651},
  year={2024}
}

@article{ouyang2025tosem,
  title={An empirical study of the non-determinism of chatgpt in code generation},
  author={Ouyang, Shuyin and Zhang, Jie M and Harman, Mark and Wang, Meng},
  journal={ACM Transactions on Software Engineering and Methodology},
  volume={34},
  number={2},
  pages={1--28},
  year={2025},
  publisher={ACM New York, NY}
}

@misc{atil2025arxiv,
      title={Non-Determinism of "Deterministic" LLM Settings}, 
      author={Berk Atil and Sarp Aykent and Alexa Chittams and Lisheng Fu and Rebecca J. Passonneau and Evan Radcliffe and Guru Rajan Rajagopal and Adam Sloan and Tomasz Tudrej and Ferhan Ture and Zhe Wu and Lixinyu Xu and Breck Baldwin},
      year={2025},
      eprint={2408.04667},
      archivePrefix={arXiv},
      primaryClass={cs.CL},
      url={https://arxiv.org/abs/2408.04667}, 
}

@article{fleiss1971,
  author  = {Fleiss, Joseph L.},
  title   = {Measuring Nominal Scale Agreement Among Many Raters},
  journal = {Psychological Bulletin},
  year    = {1971},
  volume  = {76},
  number  = {5},
  pages   = {378--382},
  doi     = {10.1037/h0031619},
  publisher = {American Psychological Association}
}

@article{benjamini1995,
  title={Controlling the false discovery rate: a practical and powerful approach to multiple testing},
  author={Benjamini, Yoav and Hochberg, Yosef},
  journal={Journal of the Royal statistical society: series B (Methodological)},
  volume={57},
  number={1},
  pages={289--300},
  year={1995},
  publisher={Wiley Online Library}
}

@article{landis1977,
  author  = {Landis, J. R. and Koch, G. G.},
  title   = {The Measurement of Observer Agreement for Categorical Data},
  journal = {Biometrics},
  year    = {1977},
  volume  = {33},
  number  = {1},
  pages   = {159--174},
  pmid    = {843571}
}

@article{laiq2026automatic,
  title={Automatic techniques for issue report classification: A systematic mapping study},
  author={Laiq, Muhammad and Dobslaw, Felix},
  journal={Automated Software Engineering},
  volume={33},
  number={3},
  pages={72},
  year={2026},
  publisher={Springer}
}

@article{turpin2023language,
  title={Language models don't always say what they think: Unfaithful explanations in chain-of-thought prompting},
  author={Turpin, Miles and Michael, Julian and Perez, Ethan and Bowman, Samuel},
  journal={Advances in Neural Information Processing Systems},
  volume={36},
  pages={74952--74965},
  year={2023}
}

@article{fan2026AutomaticExtrinsic,
  author  = {Fan, Guisheng and Liang, Yuguo and Zu, Longfei and Yu, Huiqun and Huang, Zijie and Chen, Wentao},
  title   = {Automatic Identification of Extrinsic Bug Reports for Just-In-Time Bug Prediction},
  journal = {Science of Computer Programming},
  year    = {2026},
  volume  = {249},
  pages   = {103410},
  doi     = {10.1016/j.scico.2025.103410},
  url     = {https://doi.org/10.1016/j.scico.2025.103410}
}

@misc{wang2023selfconsistencyimproveschainthought,
      title={Self-Consistency Improves Chain of Thought Reasoning in Language Models}, 
      author={Xuezhi Wang and Jason Wei and Dale Schuurmans and Quoc Le and Ed Chi and Sharan Narang and Aakanksha Chowdhery and Denny Zhou},
      year={2023},
      eprint={2203.11171},
      archivePrefix={arXiv},
      primaryClass={cs.CL},
      url={https://arxiv.org/abs/2203.11171}, 
}

@misc{shi2023largelanguagemodelseasily,
      title={Large Language Models Can Be Easily Distracted by Irrelevant Context}, 
      author={Freda Shi and Xinyun Chen and Kanishka Misra and Nathan Scales and David Dohan and Ed Chi and Nathanael Schärli and Denny Zhou},
      year={2023},
      eprint={2302.00093},
      archivePrefix={arXiv},
      primaryClass={cs.CL},
      url={https://arxiv.org/abs/2302.00093}, 
}

@misc{oskooei2025manyshotpromptingfailsempirical,
      title={When Many-Shot Prompting Fails: An Empirical Study of LLM Code Translation}, 
      author={Amirkia Rafiei Oskooei and Kaan Baturalp Cosdan and Husamettin Isiktas and Mehmet S. Aktas},
      year={2025},
      eprint={2510.16809},
      archivePrefix={arXiv},
      primaryClass={cs.SE},
      url={https://arxiv.org/abs/2510.16809}, 
}

@inproceedings{JiangNam2026CursorRules,
  author    = {Jiang, Shaokang and Nam, Daye},
  title     = {Beyond the Prompt: An Empirical Study of Cursor Rules},
  booktitle = {Proceedings of the 23rd International Conference on Mining Software Repositories},
  year      = {2026},
  note      = {To appear},
  doi       = {10.1145/3793302.3793367},
  url       = {https://arxiv.org/abs/2512.18925},
}

@article{kallis2021predicting,
  title={Predicting issue types on GitHub},
  author={Kallis, Rafael and Di Sorbo, Andrea and Canfora, Gerardo and Panichella, Sebastiano},
  journal={Science of Computer Programming},
  volume={205},
  pages={102598},
  year={2021},
  publisher={Elsevier}
}

@inproceedings{kallis2019ticket,
  title={Ticket tagger: Machine learning driven issue classification},
  author={Kallis, Rafael and Di Sorbo, Andrea and Canfora, Gerardo and Panichella, Sebastiano},
  booktitle={2019 IEEE International Conference on Software Maintenance and Evolution (ICSME)},
  pages={406--409},
  year={2019},
  organization={IEEE}
}

@inproceedings{aracena2024applying,
  title={Applying large language models to issue classification},
  author={Aracena, Gabriel and Luster, Kyle and Santos, Fabio and Steinmacher, Igor and Gerosa, Marco Aurelio},
  booktitle={Proceedings of the Third ACM/IEEE International Workshop on NL-based Software Engineering},
  pages={57--60},
  year={2024}
}

@article{wu2021data,
  title={Data quality matters: A case study on data label correctness for security bug report prediction},
  author={Wu, Xiaoxue and Zheng, Wei and Xia, Xin and Lo, David},
  journal={IEEE Transactions on Software Engineering},
  volume={48},
  number={7},
  pages={2541--2556},
  year={2021},
  publisher={IEEE}
}

@article{fan2018chaff,
  title={Chaff from the wheat: Characterizing and determining valid bug reports},
  author={Fan, Yuanrui and Xia, Xin and Lo, David and Hassan, Ahmed E},
  journal={IEEE transactions on software engineering},
  volume={46},
  number={5},
  pages={495--525},
  year={2018},
  publisher={IEEE}
}

@article{colavito2026issue,
  title={Issue Classification with LLMs: An Empirical Study of the NASA flight Software Systems},
  author={Colavito, Giuseppe and Lanubile, Filippo and Novielli, Nicole and Arreza, Christopher and Shi, Ying},
  journal={Journal of Systems and Software},
  pages={112851},
  year={2026},
  publisher={Elsevier}
}

@article{zhang2025less,
  title={Less is more: On the importance of data quality for unit test generation},
  author={Zhang, Junwei and Hu, Xing and Gao, Shan and Xia, Xin and Lo, David and Li, Shanping},
  journal={Proceedings of the ACM on Software Engineering},
  volume={2},
  number={FSE},
  pages={1293--1316},
  year={2025},
  publisher={ACM New York, NY, USA}
}

@article{khot2022decomposed,
  title={Decomposed prompting: A modular approach for solving complex tasks},
  author={Khot, Tushar and Trivedi, Harsh and Finlayson, Matthew and Fu, Yao and Richardson, Kyle and Clark, Peter and Sabharwal, Ashish},
  journal={arXiv preprint arXiv:2210.02406},
  year={2022}
}

\end{document}